\documentclass[12pt]{iopart}

\usepackage{iopams}
\usepackage[dvips]{graphicx}
\usepackage{epsfig}

\begin{document}

\title[Memory-based foraging and plant ecosystem self-organization]
{Self-organization, scaling and collapse in a coupled automaton model 
of foragers and vegetation resources with seed dispersal} 

\author{D Boyer and O L\'opez-Corona}

\address{Instituto de F\'\i sica, Universidad
Nacional Aut\'onoma de M\'exico, Apartado Postal 20-364, 
01000 M\'exico D.F., M\'exico}

\ead{\mailto{boyer@fisica.unam.mx}}

\begin{abstract}
We introduce a model of traveling agents ({\it e.g.} frugivorous animals) 
who feed on randomly located vegetation patches and disperse their 
seeds, thus modifying the spatial distribution of resources in the long term.
It is assumed that the survival probability of a seed increases with the 
distance to the parent patch and decreases with the size of the colonized 
patch. In turn, the foraging agents use a deterministic strategy with memory,
that makes them visit the largest possible patches accessible within minimal 
travelling distances. The combination of these interactions produce complex 
spatio-temporal patterns. If the patches have a small initial size, 
the vegetation total mass (biomass) increases with time and reaches a 
maximum corresponding to a self-organized critical state with power-law 
distributed patch sizes and L\'evy-like movement patterns for the foragers. 
However, this state collapses as the biomass sharply decreases to reach 
a noisy stationary regime characterized by corrections to scaling. In systems 
with low plant competition, the efficiency of the foraging rules leads to
the formation of
heterogeneous vegetation patterns with $1/f^{\alpha}$ frequency spectra, 
and contributes, rather counter-intuitively, to lower the biomass levels.
\end{abstract}

\submitto{\JPA}
%\maketitle

\section{Introduction}

Animal movement and its ecological implications is a discipline that attracts 
a growing interest \cite{nathanpnas}. The study of animal displacements
gives valuable clues on how complex organisms adapt to their environment,
in particular to search, prepare and consume food \cite{stevenskrebs}. 
Foraging problems have motivated many modeling approaches, 
sometimes inspired from the physics of the random walk problem
\cite{turchin}. 
In a way similar to anomalously diffusing
particles in a physical context, the displacement patterns of a variety of
animals (albatrosses \cite{vis}, bumble-bees \cite{vis}, primates \cite{gabriel},
gastropods \cite{gastro}, jackals \cite{atk}, seals \cite{seals} and sharks
\cite{sims}, among others) involve many spatio-temporal scales and are 
sometimes 
well described by L\'evy walks or intermittent processes with power-law
distributions. On a biological point of view, wide fluctuations in 
the movements of a herbivorous or frugivorous animal are interesting as they 
may reflect a variety of behavioural responses induced by a complex 
environment with resources distributed heterogeneously 
\cite{hasselmay,bovetbenhamou,benhamou2007,boyer2006}.

Plant ecosystems are out-of-equilibrium and exhibit
rich structures and dynamics. Spatial patterns in the
distribution of plant species are highly non random and contain
many characteristic length scales \cite{condit2000}. Patch 
distributions in rain forests have fractal properties, suggesting 
that these systems could be near a self-organized critical state driven 
by the slow growth of trees and sudden mortality avalanches \cite{solemanrubia}.
Other observations report that
tree sizes (and therefore fruit contents) in template and tropical forests 
are distributed according to inverse power-laws \cite{enquist,niklas}. When water 
resources are a limiting factor, 
continuous models show that plant interactions produce aggregation in patches 
that self-organize at larger scales to form more or less regular 
patterns \cite{vegpatternsprl}, or disordered ones where patch areas
obey scale-free probability distribution functions \cite{scalefreeveg}.

Seed dispersal represents an important animal/plant interaction that may 
contribute to the formation of complex ecological patterns. 
Seed dispersal at long distances has been identified as an important 
structuring factor of tree communities \cite{janzen,nathan}. Fruit eating 
animals ({\it e.g.}, spider monkeys \cite{lambert}) swallow the seeds 
of many tree species and deposit them, through faeces, practically intact 
and away from the parent tree after a transit time of a few hours. 
Between 60 and 90$\%$ of the seeds of tree species of tropical forests are 
dispersed by vertebrates that feed on fruit \cite{jordano}, specially 
primates \cite{clarkpoulsen,wehncke2003,russo}.

The aim of this article is to study an automaton model of moving foragers 
that modify, via seed dispersal, the long term structure of the resources
they consume. In turn, these resources also determine the foragers 
displacements, who use cognitive skills to explore their medium in an 
efficient, non-random way. The model assumes that two factors influence 
the growth success of a seed: the distance to the parent plant \cite{janzen} 
and competition due to the presence of other plants \cite{nathan}.
Despite that the model is over-simplified, the forager/resources coupled 
dynamics leads to rich behaviours, like self-organized states with 
power-law statistics for patch sizes and animal movement lengths.

The following two Sections describe the model and its background.
The results and the underlying mechanism leading to power-laws in this system 
are presented in Section \ref{res}, and conclusions in Section \ref{concl}.

\section{Background: a foraging model without plant dynamics}

We first describe the movement rules of the model forager in a stationary
distribution of resource patches (see refs. \cite{boyer2006,boyer2009}).
Consider a two-dimensional square domain of area unity containing $N$ 
fixed, point-like patches randomly and uniformly distributed. To each patch 
$i$ is assigned a fixed size (or food content), $k_i=1,2...$, a integer 
drawn from a given distribution, $p(k)$.

Like many other animals, primates use cognitive maps to navigate 
their environment \cite{garber1989,janson1998}. Evidence shows that
travels to fruiting trees are more frequent than as suggested by random 
null models \cite{janson1998}. Primates also keep record of the sites they have 
visited in a recent past \cite{garber1989}.
For simplicity, we assume that our model forager has a perfect knowledge of
the positions and sizes of all the patches in the system. Initially, a 
forager is located on a patch chosen at random. The following
deterministic foraging rules are then iteratively applied at every time step
($t\rightarrow t+1$):

\hspace{1cm}{\it (i)} The forager located at patch $i$ feeds on that patch, 
the fruit content decreasing by one unit: $k_i\rightarrow k_i-1$.

\hspace{1cm}{\it (ii)} If $k_i$ has reached the value 0, the forager chooses an 
other patch, $j$, such that $k_j/d_{ij}$ is maximal over all the allowed 
patches $j\neq i$ in the system, where $k_j$ is the food content of patch $j$ 
and $d_{ij}$ the Euclidean distance between patches $i$ and $j$. With this rule, 
the next visited patch (the \lq\lq best" patch) has a large food content 
and/or is at a short distance from $i$. We assume that 
the travel from $i$ to $j$ takes one time unit. 

\hspace{1cm}{\it (iii)} The forager does not revisit previously visited patches. 

This model produces complex trajectories that have been studied 
in details in refs. \cite{boyer2006,boyer2009} and discussed in connection with
spider monkeys foraging patterns observed in the field \cite{gabriel}. 

The model has a remarkable property, of interest in the following. 
Let us define the forager
mean-displacement $R(t)$ as $\langle|{\bf r}(t+t_0)-{\bf r}(t_0)|\rangle$
with ${\bf r}(t)$ the forager position at time $t$, the averages
being taken over different times $t_0$ and independent disorder 
realizations. At fixed patch number $N$ and time $t$, if the resource size 
distribution is the inverse power-law $p(k)=ck^{-\beta}$ with $\beta=3$, then 
the mean displacement $R(t)$ is maximal \cite{boyer2006}.
In other words, media with this size distribution induce maximal 
displacements, see figure 1. (This property still holds if the forager travels 
at constant velocity instead of moving in one time unit from one patch to the 
other \cite{boyer2006}.)

The feature above can be understood qualitatively by noting that
if the medium is very homogeneous (say, $\beta\gg 1$), 
then all patches are similar in size: given rule {\it (ii)}, the forager 
chooses essentially nearby patches. Trajectories
are thus composed of small steps and diffusion is relatively slow 
\cite{boyer2006}. On the contrary, if the medium is very heterogeneous 
($\beta\simeq 1$), patches with $k_i\gg 1$ are numerous: the forager 
often performs a large step to reach a very good patch and stays there 
a long time feeding, given rule ${\it (i)}$. The forager activity is 
dominated by these long trapping times, resulting in very slow (nearly frozen)
diffusion. An intermediate situation corresponds to $\beta=3$, for which
the best patch from a given point is often far away (at distances much
larger than the typical distance between nearest-neighbour patches), but these
good patches still have reasonable sizes, so that the forager does not 
remain trapped feeding on them during very long periods of time.

\begin{figure}
\centering \includegraphics[width=0.7\textwidth]{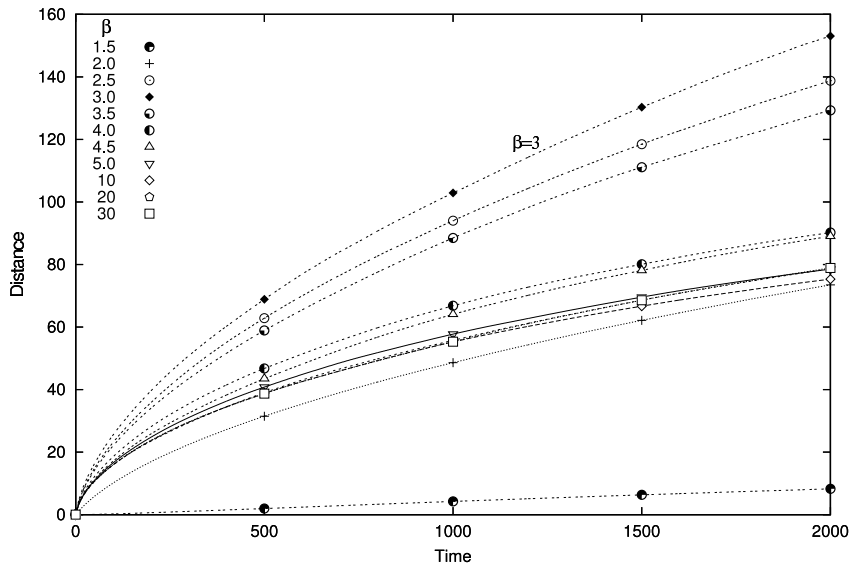}

\vspace{0.5cm}
\caption{Numerically obtained mean displacement (in unit of $l_0=N^{-1/2}$) 
as a function of time for the model forager in a random medium with resources 
size distributed as $p(k)=C k^{-\beta}$ ($N=10^6$).}
\label{figdispersion}
\end{figure}

Let us note that, at the special resource exponent value $\beta=3$, 
the trajectory of the forager closely resembles a L\'evy flight. 
Numerical simulations show that the distribution function of the distances 
separating successively visited patches is asymptotically given
by the power-law \cite{boyer2006,boyer2009}
\begin{equation}\label{scalelength}
P(l)\sim l^{-2}.
\end{equation}
The foraging patterns of spider monkeys are well described by the distribution 
(\ref{scalelength}) \cite{gabriel,boyer2006}. Even more, these animals feed on 
trees whose size distribution obeys $p(k)\sim k^{-\beta}$ with 
$\beta\simeq2.6$, a value close to 3 \cite{boyer2006}.

\section{Modified model with plant dynamics}

In the previous model, forager motion is induced by the medium. In the generalization
considered from now on, the forager follows the same rules but also modifies 
its environment through seed dispersal. Hence, the distribution of resource sizes, 
$p(k)$, is no longer held fixed and can slowly change over time. We assume that
foragers are the main mechanism of seed dispersal.

At $t=0$, $N$ point-like patches, all with size $k_i=1$, are randomly and uniformly 
distributed in the square domain of unit area. A forager initially located on a patch
chosen at random follows the rules {\it (i)}-{\it (iii)} above. In addition,

\hspace{1cm}{\it (iv)} every $\tau_{d}$ time units (the digestion time), a 
seed is deposited at the patch where the forager is located;

\hspace{1cm}{\it (v)} every $\tau_{walk}(\gg\tau_d)$ time units, 
the walk ends and the forager is removed; the patches are refreshed to their
initial $k_i$ values; the patches that have received a seed 
that has survived (see below) increase their size by one unit, 
$k_i\rightarrow k_i+1$;

\hspace{1cm}{\it (vi)} a new forager ({\it i.e.} not representing 
necessarily the same individual, but still having a perfect knowledge of 
the updated environment) is located on a patch chosen at random and the process
is iterated as above for another $\tau_{walk}$ time units;

\hspace{1cm}{\it (vii)} a plant that has grown from a seed deposited $\tau_m$
time units earlier ($\tau_m\gg\tau_{walk}$) dies and the patch size decreases, 
$k_i\rightarrow k_i-1$;

\hspace{1cm}{\it (viii)} in rule {\it (vii)}, the size of a patch does not 
decrease below the minimal value $k_i=1$.

We assume that two factors determine the survival probability of a deposited seed
(or the growth success of a plant here) in stage {\it (v)}.
The first assumption is based on observations that seeds dispersed far away
survive better, as they are less likely to attract seed predators and to 
be transmitted parasites or diseases from the parent plant \cite{janzen}. 
The second assumption takes into account competition among plants for limited
nutrients \cite{nathan}. Let us note $k$ the size of the patch where the seed
is deposited and $l$ its distance to the parent patch, which is the
patch where the forager was located $\tau_d$ time units ago. 
The survival probability of the seed, $P_s$, is set to $P_s=P_d(l)P_c(k)$ with
\begin{eqnarray}
P_d(l)&=&\left[l/(l+l_0)\right]^n,\label{Pd}\\ 
P_c(k)&=&\left\{
\begin{array}{ll}
1-\frac{k}{k_{max}+1}\ &{\rm if}\ k\le k_{max}\\
0\ &{\rm if}\ k> k_{max},
\end{array}
\right.\label{Pc}
\end{eqnarray}
with $l_0=N^{-1/2}$ the typical distance between nearest-neighbour patches and 
$k_{max}$ a fixed integer accounting for competition effects. 
The parameter $k_{max}$ is proportional, say, to the nutrients concentration.
According to (\ref{Pd}), the survival probability increases from 0 to 1 as the 
distance from the parent patch increases, whereas it decreases from 1 to 0 as 
the size of the colonized patch increases from 1 to $k_{max}$, the maximum 
patch size. In the following, we fix $n=1$ in Eq.(\ref{Pd}), the results 
presented below being not qualitatively modified if larger values of $n$ 
are chosen.

\section{Results}\label{res}

It is natural to ask the following question.
Given that in the model the survival probability increases with the distance 
to the parent plant and that the largest forager displacements are produced
in media with the size distribution $p(k)\sim k^{-3}$, do resources 
self-organize towards this particular scaling-law (and foraging patterns
towards the L\'evy law (\ref{scalelength}))? As a consequence of the memory-based
foraging rules {\it (i)-(iii)}, such environments should have the highest average 
biomass production rate. As the foragers start dispersing seeds, small heterogeneities 
are produced ($k_i\neq 1$), and, from a dynamical system point of view, the fastest 
growing modes may dominate the dynamics asymptotically.

\begin{figure}
\begin{center}
\includegraphics[width=0.7\textwidth]{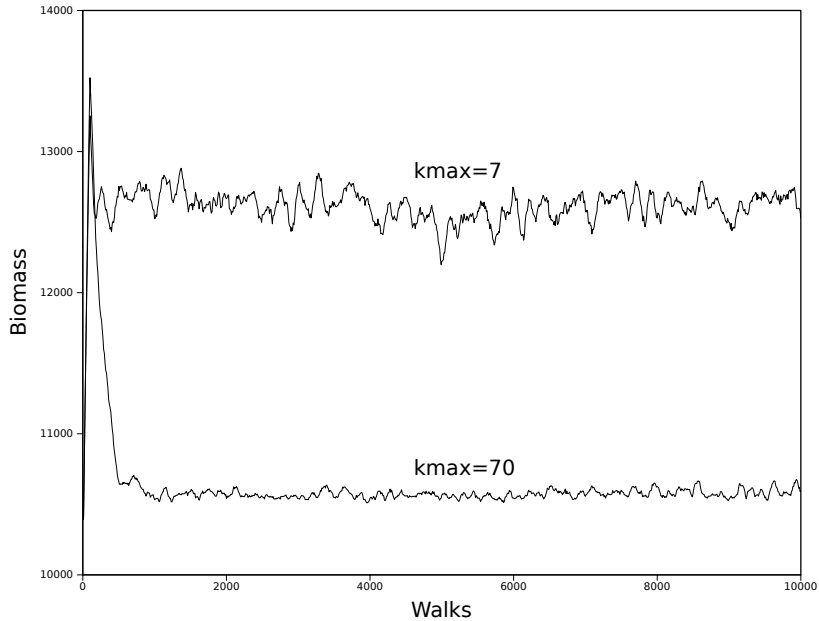} 
\end{center}
\caption{Time series (in units of $\tau_{walk}$) of the biomass,
for two values of the competition parameter $k_{max}$. The
parameters are $N=10^4$, $\tau_{d}=10$, $\tau_{walk}=500$ and
$\tau_m=5\ 10^4$.}
\label{figbiomass}
\end{figure}

Successful seeds contribute to the emergence of larger, more attractive patches 
that are also susceptible of being visited more often in the future.
This positive feedback loop contributes to the increase of heterogeneities,
within some limits. If very big patches are produced, due to rule 
{\it (i)} many seeds will be dispersed at the distance
$l=0$ and will die, given the kernel (\ref{Pd}). The other stabilizing effects 
are competition and mortality.

\subsection{The rise and fall of scaling}

To investigate the possibility of the scenario sketched above, let us first
consider the behaviour of the system biomass, $M=\sum_{i=1}^N k_i$, 
as a function of time. As displayed in figure \ref{figbiomass}, $M$ initially
increases and reaches a maximum value. It then suffers a abrupt drop,
followed by a noisy stationary regime. The asymptotic average biomass depends
strongly on plant competition: unexpectedly, it is lower at low competition levels
(large $k_{max}$). 

\begin{figure}
\hspace{1.5cm} \epsfig{figure=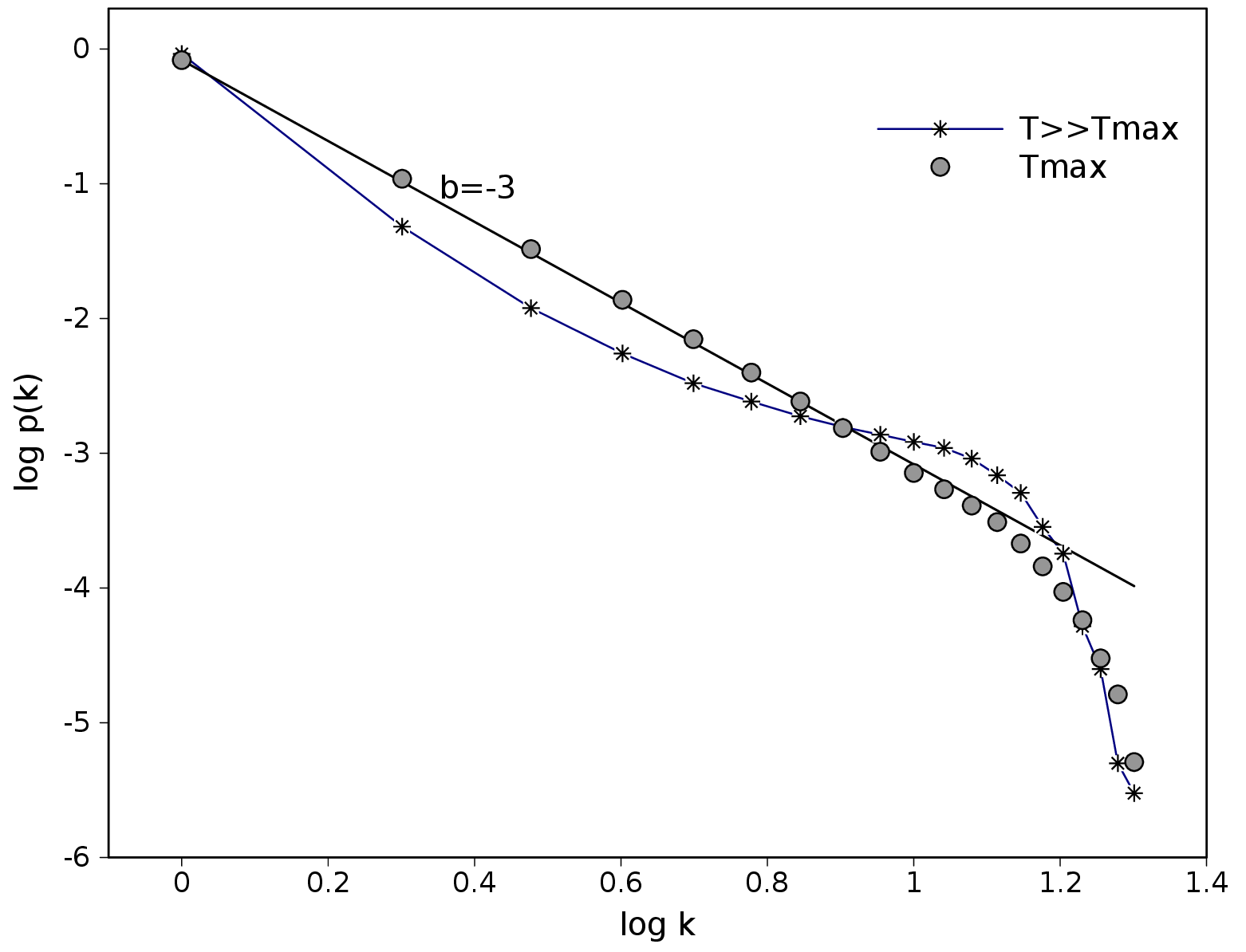,width=2.8in}
\epsfig{figure=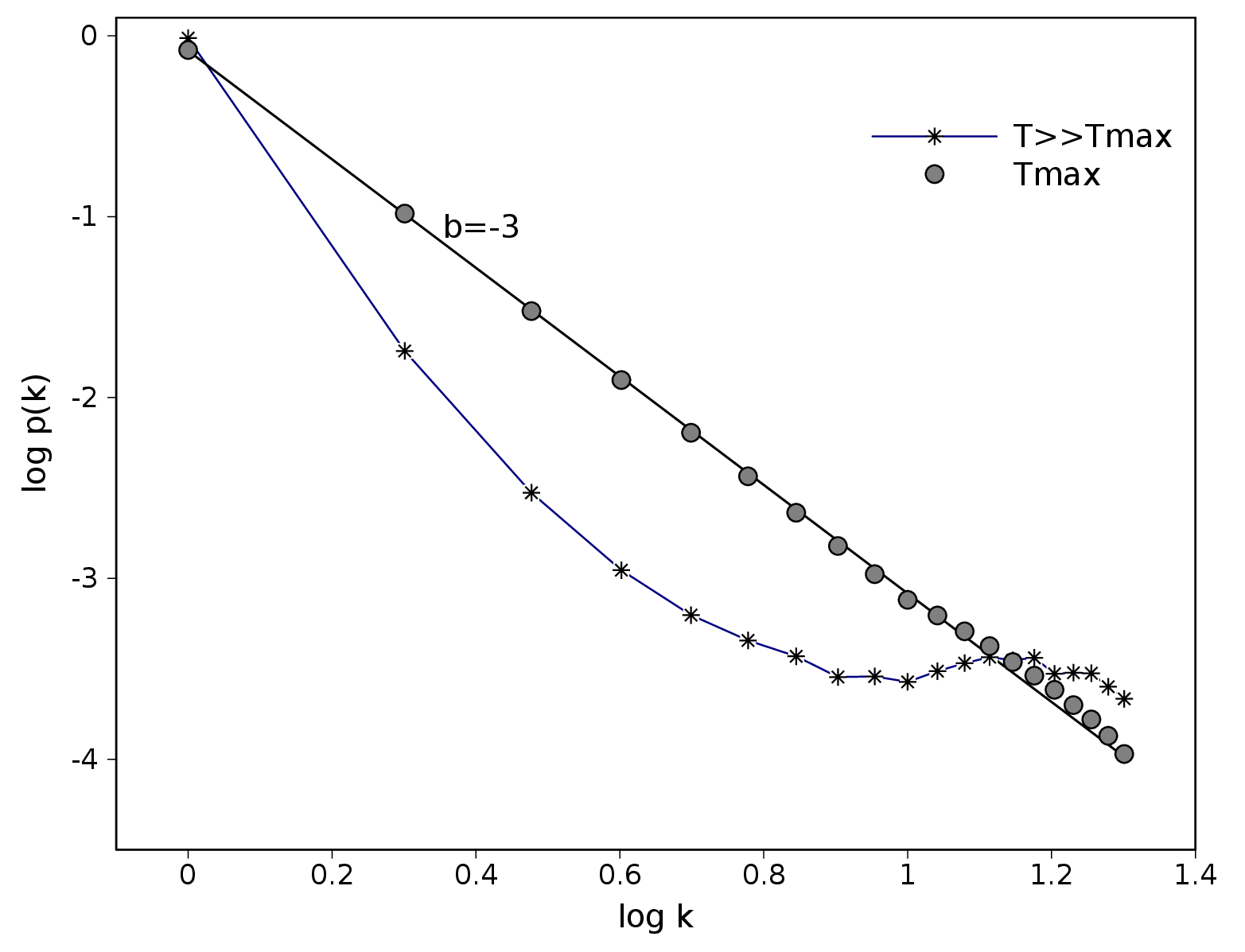,width=2.8in}
\caption{Patch size distributions (in log-decimal representation) 
for $k_{max}=20$ (left) and $k_{max}=50$ (right), at two different 
times: $t_{max}(=100\tau_{walk},$ here), when the biomass is maximum 
(see fig.\ref{figbiomass}), and $t=5000\tau_{walk}$, in the stationary 
regime. Averages are performed over $10^3$ independent initial conditions.
The straight lines have a slope $-3$.}
\label{figpk1}
\end{figure}

\begin{figure}
\centering \includegraphics[width=0.6\textwidth]{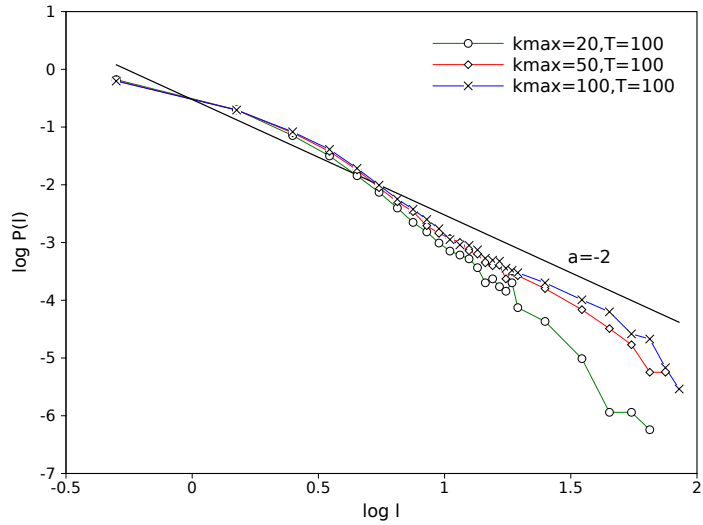}
\caption{Forager step length distributions at $t_{max}$, when biomass
is maximal, for $k_{max}=20,50$ and $100$. The straight line has a slope 
$-2$.}
\label{figpl}
\end{figure}

The system actually builds spatial heterogeneities during the
initial growth regime. Biomass is maximum at $t_{max}=\tau_m=(100\tau_{walk}$,
here), when the first plants grown from dispersed seeds start to die.
As shown by figure \ref{figpk1}, the size distribution $p(k)$ at 
$t=\tau_{m}$ is perfectly fitted by the power-law $k^{-3}$ for $k\ll k_{max}$, 
independently of the value of $k_{max}$. At the same time, and as expected from 
Eq.(\ref{scalelength}), the forager step length distribution (figure \ref{figpl}) 
tends to the scaling-law $P(l)\sim l^{-2}$ in a range of intermediate values of $l$, 
with a truncation at large $l$ due to finite $k_{max}$, $\tau_m$ and $N$.
Note the existence of steps of order of the system size ($100l_0$, here). 
In some sense, the system self-organizes into a critical state of maximal 
dispersion.

Interestingly, this scaling behaviour does not persist at larger times 
$t\gg \tau_m$. Simultaneously to the biomass collapse (Fig.\ref{figbiomass}), 
corrections to scaling appear. Asymptotically, the shape of the patch 
distribution decreases rapidly with patch size and exhibits a practically 
constant fat tail, see figure \ref{figpk1}. Hence, most patches have a small 
characteristic size and coexist with a few \lq\lq outliers" of size of order 
$k_{max}$. This separation into two characteristic sizes becomes more 
pronounced at large $k_{max}$.

Unexpectedly, dispersal by animals does not manage to stabilize the 
critical state for which it is responsible. Despite that the system rapidly 
self-organizes into a state with optimal seed survival, it becomes 
overpopulated: mortality is much higher than birth shortly after $\tau_m$ 
and the size of many intermediate patches start to shrink. This feature is 
probably due to the fact that, at $t=\tau_m$, the system is much more 
crowded with plants than at $t=0$. The birth rate is therefore lower 
than in the initial growth regime because of higher competition (Eq.(\ref{Pc})):
older dying plants are not replaced by the same quantity of new ones.
In addition, due to their cognitive maps and the rules {\it(i)-(iii)}, 
foragers neglect many small, unattractive patches, among them the shrinking 
ones, that would be suitable for plant growth. Instead, they keep revisiting 
a small number of nearly saturated large patches (outliers).

\subsection{Biomass in the stationary regime}

\begin{figure}
\hspace{-0.4cm} \epsfig{figure=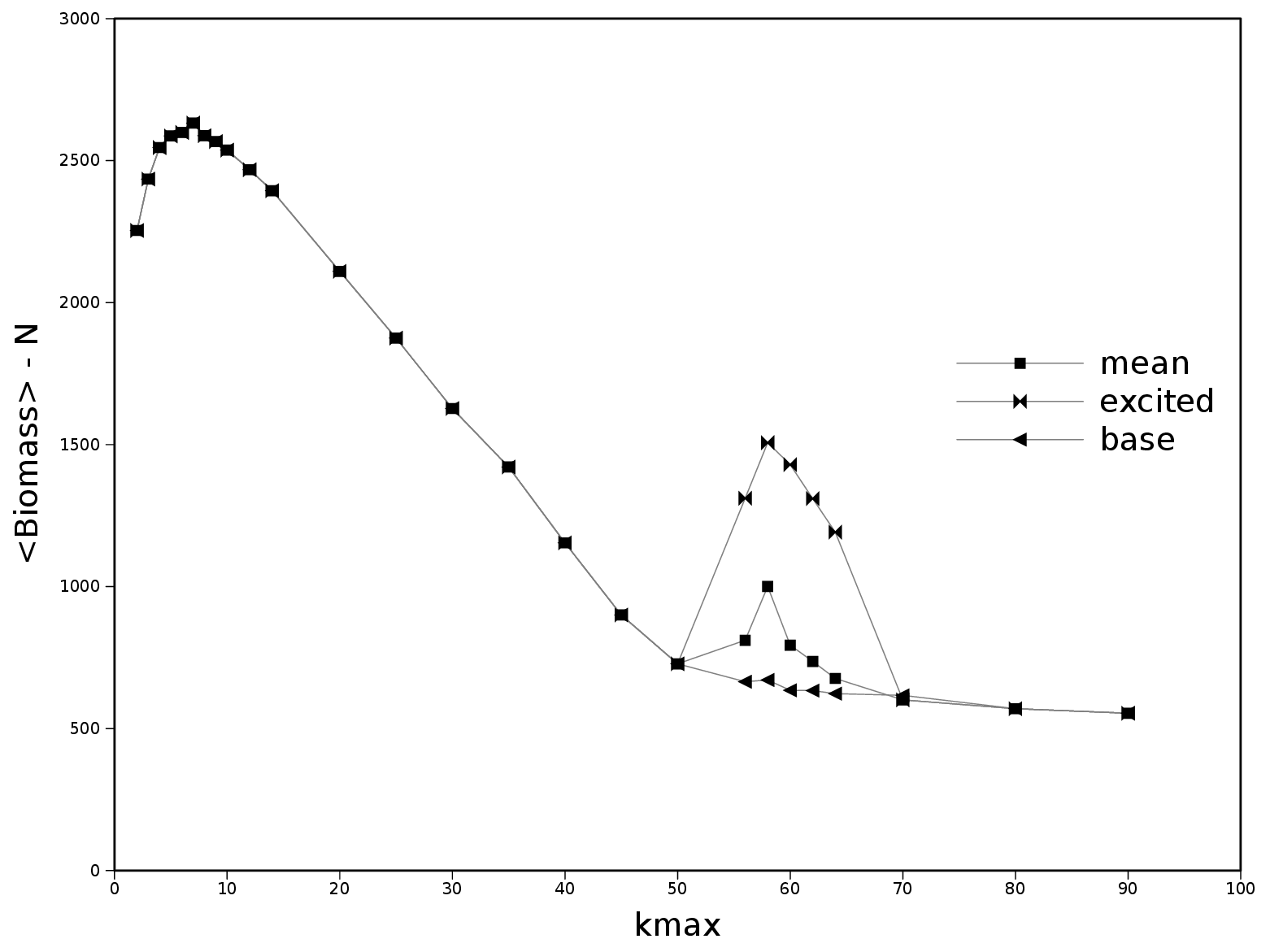,width=3.1in}
\epsfig{figure=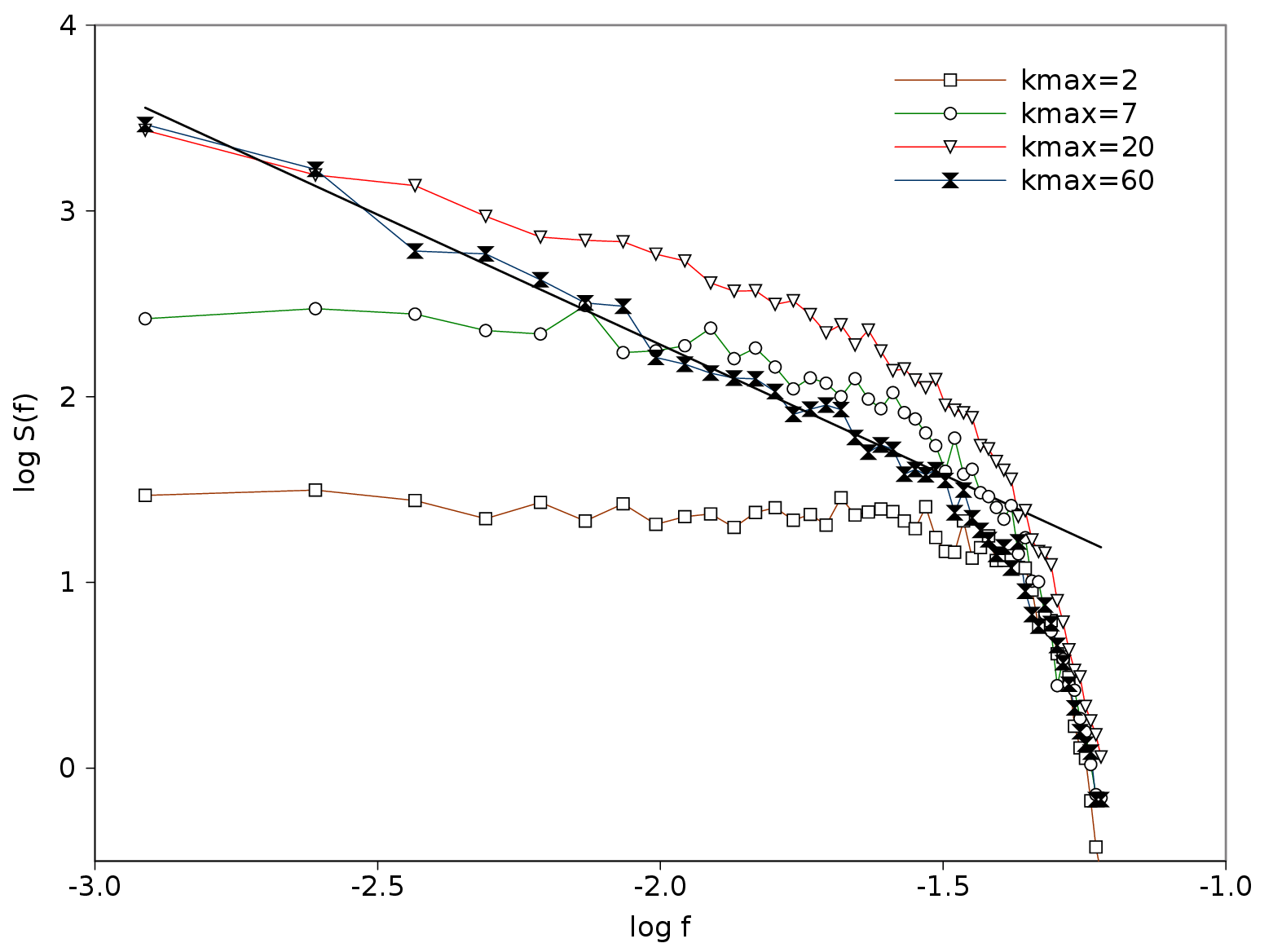,width=3.1in}
\caption{{\it Left}: Mean value of the biomass in the stationary regime
for several values of the parameter $k_{max}$. Averages are performed 
over 100 independent runs. The curve has a maximum at $k_{max}=7$ 
(corresponding to a fairly strong plant competition) and a bistable 
region centered around $k_{max}=58$ (low plant competition). Same parameters
as in Fig. \ref{figbiomass}. {\it Right}: Log-decimal power 
spectrum $S(f)\equiv\langle \tilde{M}(f)\tilde{M}(-f)\rangle$, 
with $\tilde{M}(f)$ the Fourier transform of the biomass $M(t)$ in the 
stationary regime (see figure \ref{figbiomass}), for $k_{max}=2,7,20$ 
and $60$. Averages are taken over 60 runs. The straight line is a power-law 
with exponent $-1.4$.}
\label{figbiomassprom}
\end{figure}

Asymptotically, foragers therefore concentrate their activity on a fraction
of the available land, resulting in a reduced biomass. Figure 
\ref{figbiomassprom} (left) displays the average biomass in the stationary 
regime as a function of the maximum allowed patch size, $k_{max}$ (or 
nutrients concentration). Counter-intuitively, but in agreement with 
the comments above, the general tendency is a biomass decrease with 
increasing $k_{max}$. The biomass is large at small $k_{max}$ (relatively
low nutrients/high competition regime) and presents an extrema at $k_{max}=7$. 
At this parameter 
value, patches are relatively homogeneous in size: via foragers
visits, seeds can colonize many different patches. At higher $k_{max}$, 
foragers start to visit large patches preferentially, neglecting smaller ones.

Supporting this interpretation, the power-spectra of the biomass 
time series $M(t)$ reveal an increase in complexity with increasing 
$k_{max}$, see figure \ref{figbiomassprom} (right). At high competition 
levels ($k_{max}=2$), the fluctuations of $M(t)$ around its mean value 
are due to births and deaths that occur in a roughly independent way, 
due to a homogeneous dispersion of seeds. The spectrum 
is that of a white noise. At low competition ($k_{max}=60$), the power 
spectrum is well approximated by a power-law $1/f^{\alpha}$ with 
$\alpha\simeq1.4$ 
over more than one decade, indicating long-range temporal correlations. 
The spatially heterogeneous colonization of plants generates periods 
of high mortality (\lq\lq avalanches") of widely varying durations, 
followed by periods of easier recolonization. This dynamics is obviously 
reminiscent of the punctuated relaxation of sand-pile models in 
self-organized criticality \cite{bak}.

An other unexpected phenomena is observed at low competition levels, 
in the interval $50<k_{max}<70$ for the parameter values considered here. 
In this interval, the average biomass increases again and 
exhibits a second maximum at $k_{max}\simeq 58$. A closer look reveals that
the system actually converges towards two distinct states, depending 
on the initial condition. At identical parameters values, the system 
sometimes ends up in a low, \lq\lq base" biomass level that follows the 
tendency described above, and sometimes exhibits a significantly higher 
biomass (\lq\lq excited" state). Biomass fluctuations are relatively 
small among different systems belonging to a same class (base or excited), 
which makes possible the computation of the mean values separately 
(see figure \ref{figbiomassprom}). The origin of this bistability is 
unclear, although it is a known phenomenon in other models of 
vegetation pattern formation \cite{vegpatternsprl}.

\section{Discussion and conclusions}\label{concl}

We have proposed a new mechanism leading to the emergence of many 
spatio-temporal 
scales in the movement patterns of foraging animals. In the model proposed, 
foragers use mental maps to choose feeding patches and disperse seeds 
along their trajectories, thus affecting the long term distribution of their
food resources. This simplified model focuses on plant-forager interactions, 
neglecting other important factors (wind, gravity...) of seed 
dispersal \cite{nathan}. It is built on a generalization of a previous model, 
where scale-free displacement patterns of knowledgeable animals emerge from 
their interaction with resources that are distributed according to an
{\it a priori} given power-law function \cite{boyer2006,boyer2009}. 
In the field, the distributions of the movements of spider monkeys and 
of the size of their fruiting trees are in good agreement with those 
given by that model for a particular parameter value, where animal 
displacements are maxima \cite{boyer2006}. 

The present approach shows that a memory-based ranging behaviour generates 
highly heterogeneous seed deposition patterns, a conclusion also reached 
in ref. \cite{russo} with the use of a spatially explicit model parameterized 
with field-collected spider monkeys movement data. These findings suggest 
that, over large temporal scales, tree distributions may form complex 
spatial structures due to the presence of foraging animals. The present 
model proposes a theoretical test of this hypothesis: the distribution 
of resources is not held fixed and spatial heterogeneities self-organize 
spontaneously under the influence of positive feedback loops in the system 
dynamics. 

Other existing theories of L\'evy \cite{vis} or intermittent 
\cite{benichou,oshanin} foraging assume that animals are memoryless and 
do not have any information on the location of resource patches. In these 
contexts, they execute a given Markovian stochastic processes to find preys 
(usually randomly distributed in space) that are detectable only at short 
distance. Movements with nontrivial distributions or rules are optimal for 
finding preys most efficiently in some cases. Whereas this approach can 
be justified for marine animals foraging in unpredictable 
environments \cite{sims}, frugivorous vertebrates rely on fixed resources 
and memory plays an essential role \cite{garber1989,janson1998}. Nevertheless, 
the introduction of limited knowledge and the use of search modes would 
improve the realism of the modelling approach presented here.

Despite that, at each step, our model foragers maximize an efficiency function,
their foraging activity can not be considered as optimal in the long term.
The hypothesized relationship between long distance dispersal and 
species diversity in plant communities \cite{boyer2006} is probably 
not a simple one \cite{france}.
The model exhibits sudden plant mortality avalanches that are consequences of 
a restricted land use and a lack of colonization of regions with 
low plant density. Counter-intuitively, biomass levels are lower when
the conditions for colonization are favourable, {\it i.e.} at low plant
competition levels. The same mechanism leading to the emergence of the
self-organized critical state with optimal seed dispersal is also 
responsible for its rapid collapse. This ecological \lq\lq crash" is the 
product of an intelligent foraging behaviour based on the satisfaction of 
immediate needs (feeding). Here, foragers do not change their strategy 
when plant resources start to shrink. It would be interesting to investigate
whether the introduction of noise (or \lq\lq irrationality") in the 
deterministic decision rules improves this situation.

The model developed here is similar in some aspects to sand-pile models
of self-organized criticality (SOC) \cite{bak}. The system is driven by 
successful dispersed seeds and biomass dissipated by mortality. 
SOC-like models have been proposed to explain large extinctions in the fossil 
record \cite{baksneppen} and tree dynamics on ecological time-scales in rain 
forests \cite{solemanrubia}. In the latter example, $1/f$-noise signals have 
been identified in simulated biomass time series \cite{solemanrubia}
(unfortunately, biomass in real forests is very difficult to measure). 
Qualitatively similar results are obtained here with other 
assumptions. The stationary regime of our model,
however, is not strictly speaking asymptotically critical since 
the shape of the patch size distribution contains two characteristic sizes: 
a small one, of order 1, and a much larger one, corresponding to the 
presence of \lq\lq outliers". Such distributions seem to be ubiquitous 
in driven self-organized systems, though: they appear in earthquake 
models \cite{sornettegil} and also in other models of plant dynamics 
in the presence of limited nutrients \cite{scalefreeveg}.

The robustness of the features described above should be tested
by modifying the foraging rules and the dispersion kernels. A more 
realistic, resource-dependent forager demography is an aspect that 
should also be considered.

%Many possible improvement: 
%Analytical results?
%forager mortality/natality correlated to 
%resource availability. Poissonnian seed dispersal, not deterministic. 
%Exponential increase of survival probability, instead of power law.
%Exponential decay of survival probabilitywith patch size, instead of linear.
%Growth of plant (not directly for 0 to 1). 
%Or: same model with larger taud.

\ack

This work was supported by CONACYT grant 40867-F. Fruitful discussions with 
O. Miramontes, E. Ram\'irez and P. Padilla are gratefully acknowledged. We thank 
S. Mendoza and H. Ramos for technical support.

\end{document}